# A Post-Processing Tool and Feasibility Study for Three-Dimensional Imaging with Electrical Impedance Tomography During Deep Brain Stimulation Surgery

**Sébastien Martin**[1]

[1] National Yang Ming Chiao Tung University, 1001 University Rd., Hsinchu 30010, Taiwan R.O.C.

Email: sebastien1606.eed01g@nctu.edu.tw

**Abstract**. Electrical impedance tomography (EIT) is a promising technique for biomedical imaging. The strength of EIT is its ability to reconstruct images of the body's internal structures through radiation-safe techniques. EIT is regarded as safe for patients' health, and it is currently being actively researched. This paper investigates the application of EIT during deep brain stimulation (DBS) surgery as a means to identify targets during operations. DBS involves a surgical procedure in which a lead or electrode array is implanted in a specific target area in the brain. Electrical stimulations are then used to modulate neural circuits within the target area to reduce disabling neurological symptoms. The main difficulty in performing DBS surgery is to accurately position the lead in the target area before commencing the treatment. Brain tissue shifts during DBS surgery can be as large as the target size when compared with the pre-operative magnetic resonance imaging (MRI) or computed tomography (CT) images. To address this problem, a solution based on open-domain EIT to reconstruct images surrounding the probe during DBS surgery is proposed. Data acquisition and image reconstruction were performed, and artificial intelligence was applied to enhance the resulting images. The results showed that the proposed method is rapid, produces valuable high-quality images, and constitutes a first step towards in-vivo study. Source code: https://github.com/SebastienLinker/EIT_post_processing

## 1. Introduction

Electrical impedance tomography (EIT) is a modern technology that has been proposed for low-cost biomedical imaging applications [1]. The basic idea behind EIT is to use a pair of electrodes to send an electrical current into the body, and then to measure the voltages with other electrodes located around the body. By applying an electrical current at the boundaries of the medium, it is possible to obtain the internal distribution of electrical conductivities by measuring the voltages at those boundaries, thereby solving an inverse problem [2]. Solving this inverse problem provides important information about the tissues crossed by the electrical current, and it allows those tissues to be identified.

Recently, open-domain EIT has been used for biomedical applications [3] in which electrodes are attached to a medical probe, and electrical currents are used to obtain information about the tissues surrounding the probe. The algorithms used to reconstruct the images usually generate large artefacts, although the direction of the target can be seen, it is impossible to assess the distance between the probe and the target tissue.

A method for combining linear and nonlinear methods to solve the EIT inverse problem has been proposed by Martin and Choi [4]. In this method, the ANN is applied at the output of a linear reconstruction algorithm (on the conductivity distribution in the FE model).

Recently, EIT imaging of the brain activity from the surface has been done and EIT has been a successful non penetrating method to image brain activity and neuronal charge and discharge could be seen [5]. In those studies, a non-invasive method was used [6]. In this paper, we utilize an invasive system that penetrates inside the brain in order to get a higher accuracy over a specific region of the brain during surgery.

This paper reports the results from a proposed post-processing-based solution, which was applied to 3D open-domain EIT. During deep brain stimulation (DBS) surgery, a microelectrode recording (MER) probe was inserted to locate the target area, e.g., the sub-thalamic nucleus (STN) [7]. In addition to using a traditional MER probe, using EIT as an image monitoring tool during the surgery (as a part of the MER probe) could provide real-time reconstructed images surrounding the MER probe. This approach can potentially allow neurosurgeons to identify the tissues surrounding the MER probes, including targets such as the STNs, in real-time.

### 1.1. Open-domain EIT

One major concern in open-domain EIT is the density of the mesh. Although a coarse mesh can give a rapid estimate of the conductivity distribution, a fine mesh provides greater accuracy. In the application discussed in this paper, the goal was to obtain an image of the tissue surrounding a probe, which required a fine mesh around the probe. Furthermore, a large outer boundary was required to obtain higher-quality images.

Figure 1 shows a traditional DBS lead with a microelectrode at the bottom and illustrates the process of using an open-domain EIT to provide images of the STN region.

### 1.2. Deep brain stimulation

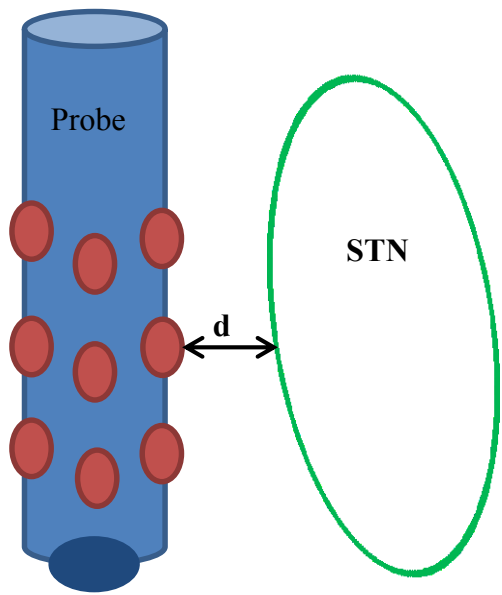

Figure 1 Illustration of the open-domain EIT problem. The DBS probe is shown in blue (with the microelectrodes shown in red), and the target, located at a distance *d* from the probe, is shown in green. The traditional microelectrode recording (MER) for measuring neuron firing is shown in deep blue.

DBS involves an advanced neurosurgical procedure that can reduce disabling neurological symptoms. To achieve this goal, a DBS lead or electrode array is inserted into a specific region of the brain. DBS modulates the electrical activity in a targeted region of the brain as a means to control debilitating neurological problems. Although the DBS procedure was first suggested in the 1950s [8], it has been practically applied only during the past 25 years. DBS aims at treating several neurological diseases such as Parkinson's disease [9], Alzheimer's disease [10], or chronic depression [11]. In this paper, we focus on Parkinson's disease, due to a degenerative STN region [9]. At present, this promising technology still faces a number of challenges [12].

One of the challenges in using this procedure is to correctly place the DBS lead, or electrode array, within the target area. Brain tissue displacements occurring during DBS surgical procedures can be comparable to the size of the targets so that the pre-planned trajectories for implantation (as determined from pre-operative magnetic resonance imaging (MRI) or computed tomography (CT) images) may no longer be accurate [13].

To solve this problem, some neurosurgeons use microelectrode recording (MER), which involves using a small electrode at the tip of a probe that monitors the firing activity of neurons surrounding the tip of the probe. MER helps to more specifically identify the precise target to be stimulated in the brain. If functional MER is used in conjunction with CT or MRI image data, it gives neurosurgeons a higher chance of accurately placing the DBS lead (or electrode array) into the target area, usually after a few trials. This MER measurement process is time-consuming, and its success is highly dependent on the neurosurgeon's skill and experience.

Although it might be harmful to stimulate the brain at this stage in the procedure [14], no harm is caused by using microelectrodes to inject low and high-frequency electrical current and then applying the EIT technique to image the surrounding region while the MER probe is being inserted. Due to this fact, it is possible to use EIT as an imaging tool that can evaluate the distance between the microelectrodes and a target region (such as the STN) in real-time during DBS surgery.

*1.3. Novelty*

The approach proposed in this paper uses artificial neural networks (ANNs) as post-processing tools to enhance the quality of EIT images by using a combination of linear and nonlinear methods. After solving the EIT inverse problem (typically through a linear solver), an ANN is used to obtain the nonlinear aspects of the conductivity distribution. By applying a direct nonlinear algorithm after using a linear method, it becomes possible to produce high-quality images in near real-time. Compared to a solution purely based on ANN, the network utilized in this application gets a significantly higher number of input data, which in turn simplifies the training of the model and reduces the risks of converging to a local minimum instead of a global minimum [15]. Unlike ANN, linear EIT inverse solvers are usually robust to the presence of noise and to miscellaneous sources of errors in the measured voltages, even in case the environment (noise, …) is not perfectly defined at training time. This proposed method is therefore robust against noise and offers a more accurate result than a method solely based on ANN.

## 2. Results

*2.1. Simulation*

The proposed method was first tested with simulations to validate the idea. In these simulations, ANNs were trained without adding any noise to the voltages. After training, different images (representing the STN region at different distances and in different orientations) were simulated by using the adjacent pattern[16], and their corresponding inverse problems were solved by using four different methods.

Cross-sections of the 3D conductivity distributions for some reconstructions are shown in Figure 2. The results from the proposed method are compared with those from the one-step Gauss-Newton (GN), the primal-dual interior point method (PDIPM) and from an ANN used as an inverse solver.

Images obtained from the linear one-step GN solver (Figure 2) show a large dislocated reconstructed object, caused by the underlying mathematics. Although this reconstruction method offers robustness to noise, it typically generates a reconstructed object that is pushed towards the outer boundary of the FE model. The resulting images show the presence and the direction of the target, but its boundaries are not clearly visible. Due to the resulting low-resolution and large artefacts, the one-step GN reconstruction method is not suitable for this application due to the distorted size of the target and significant error in distance of the reconstructed target from the probe.

However, the PDIPM algorithm is known to offer high-quality EIT reconstruction, and it can theoretically construct rough boundaries. Images obtained from this reconstruction method are shown in Figure 2 (2nd column). A target is clearly visible on the resulting images at the expected location. The reconstructed object does not seem to be pushed towards the outer boundary, and its size seems to match the expected size. The position of this reconstructed target matches the original position of the STN region, as represented by a green ellipsoid. The large ringing artefacts are visible in the reconstruction, especially around the probe and opposite the target, but the reconstructed target has rough boundaries, especially near the probe. Compared to the one-step GN, this method has the advantage of higher-quality imaging.

In closed-domain EIT, reconstruction methods based on ANNs being used as inverse solvers can produce high-quality image reconstruction. However, as can be seen in Figure 2 (3rd column), using an ANN in this configuration does not produce the expected high-quality image reconstruction. Although this solution still seems capable of imaging the STN region accurately when the target is close to the probe (Figure 2 (c)), the quality decreases drastically as the distance from the probe increases. One possible explanation for this result is that the FE model is too fine, and the ANN needs to estimate the electrical conductivity for too many elements from a limited number of measurements. Such a configuration requires both an efficient training method and efficient parameterisation of the network.

Finally, the images obtained with the proposed post-processing method are given in Figure 2 (4th column). By using this reconstruction method, the ANN was applied to the image obtained from the one-step GN method (Figure 2, 1st column), which conferred the same amounts of input and output data to the ANN. Compared to an ANN applied directly on the measurements, this method gives more information to the ANN, which in turn can take advantage of that additional information. The post-processing method then produced high-quality EIT reconstructions. A target is clearly visible in the reconstructions. In the four images obtained from this method, the location and dimensions of the target were close to the expected location and size of the original. These images show that the ringing effect (which is present in the images obtained from the one-step GN or the PDIPM) was strongly reduced by the proposed method so that this ringing effect was not visible to the human eye. In the resulting conductivity distribution, a certain quality of smoothness appeared in the artefacts as the distance between the probe and the target increased. Although this smoothness slightly degraded the visual quality of the images, the STN region remained identifiable.

Normalised average distance error (NADE), difference of resolution (|ΔRES|) and shape deformation (SD) errors were computed for the different configurations. The NADE is shown in Figure 3. The one-step GN generated a high degree of distortion of the target, and the reconstructions obtained showed a target pushed to the outer boundary of the domain. Based on the observation of these images, the NADE was expected to be high, and

it varied from 2.67 to 4.38, which confirmed the visual evidence (Figure 2). A PDIPM solver generated a visible target with rough boundaries near the probe. However, as the distance from the probe increased, parts of the boundary of the target became difficult to identify accurately. After thresholding, there was a risk that the smoothness leads to a larger object, which would therefore increase the NADE. When using a PDIPM solver, the NADE increased from 1.01 to 4.02. Although using an ANN as an inverse solver could give an accurate reconstruction when the target was close to the probe, the ANN rapidly grew more inaccurate as the distance increased, and therefore the degree of error rose rapidly. In this situation, the NADE varied from 0.53 to 2.30. The proposed post-processing scheme maintained accuracy longer than the ANN used as an inverse solver, and the level of error, therefore, remained low. Our method gave the lowest NADE, which ranged from 0.28 to 1.69.

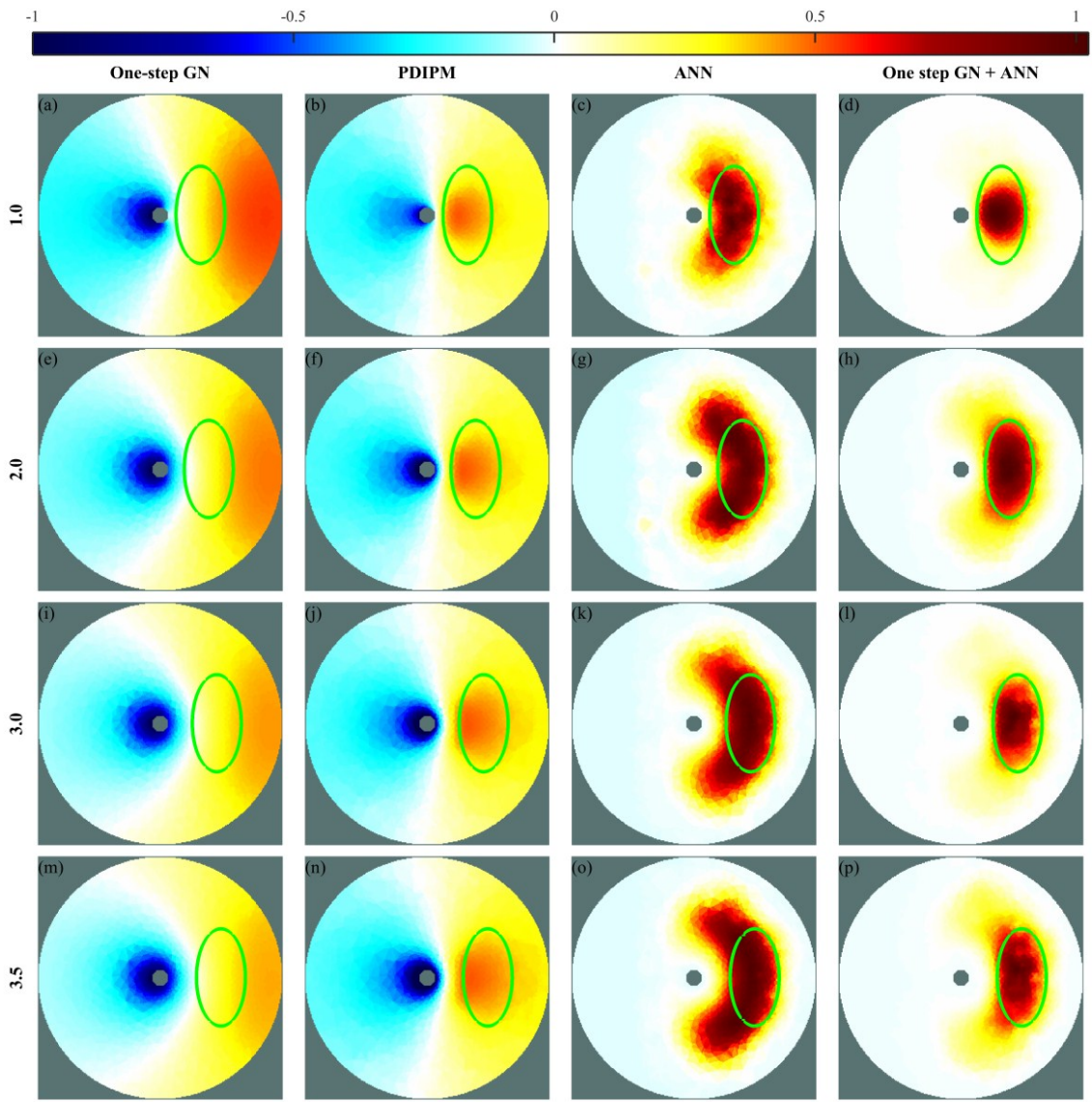

Figure 2 EIT reconstructions of the STN target, from data simulated by four different methods, namely the one-step GN (1st column), the PDIPM (2nd column), the ANN used as inverse solver (3rd column) and the proposed post-processing method (4th column). The normalised distance between the probe and the target is shown on the left. The green ellipse represents the expected location of the STN target. The normalised resistivity distribution is on top. The corresponding errors are shown in Figure 3 (NADE), Figure 4 (|ΔRES|) and Figure 5 (SD). Generated with EIDORS v3.8: http://eidors3d.sourceforge.net/

Figure 4 shows the resulting |ΔRES| errors. Again, the large degree of distortion present in the images obtained from the one-step GN resulted in a high |ΔRES| error, ranging from 17.83% to 31.40%. Although the iterative PDIPM algorithm generated high-quality images, the numbers of artefacts visible in the images substantially degraded the quality of the thresholding and therefore increased the |ΔRES| errors. The |ΔRES| errors obtained from the PDIPM solver ranged from 0.24% to 34.27%. Similarly, the ANN used as an inverse solver gave a |ΔRES| error of between 0.83% and 23.68%. The proposed post-processing method gave the lowest increase of

|ΔRES| error, of between 0.50% (when the target was close to the probe) and 19.63% (when the distance was four times the radius of the probe).

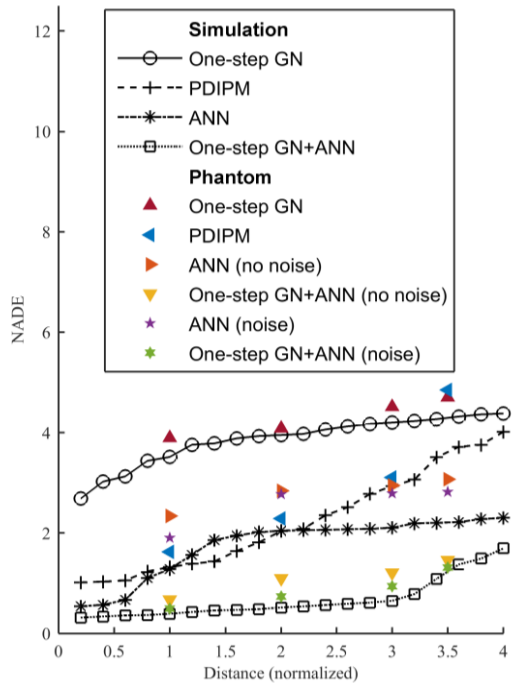

Figure 3 NADE vs. distance from probe to target, as obtained from simulation and phantom experiments using four different methods. The distance is normalised based on the radius of the probe. Detailed values are given in Supplementary Table 1 (simulations) and Supplementary Table 4 (phantom).

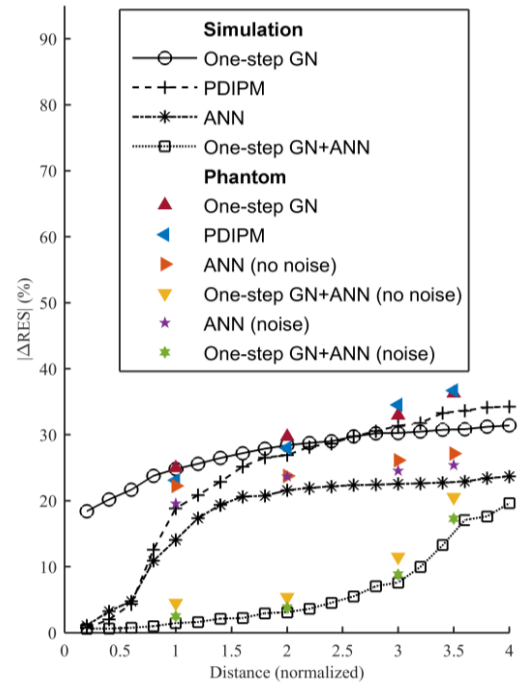

Figure 4 |ΔRES| vs. distance from probe to target, as obtained from simulation and phantom experiments using four different methods. The distance is normalised based on the radius of the probe. Detailed values are given in Supplementary Table 2 (simulations) and Supplementary Table 5 (phantom).

Like the |ΔRES| errors, the SD errors shown in Figure 5 clearly had a strong advantage for the proposed post-processing method. With simulation data, the one-step GN method always gave an SD above 66%. With the iterative PDIPM method, the resulting SD error increased from 21.43% to 67.91%. When used as an inverse solver, the ANN gave an SD error of 3.01% when the target was close to the probe, and then it quickly exceeded 20%, reaching a maximum of 57.25%. Finally, the lowest increase was obtained by applying an ANN to the conductivity distribution obtained from the one-step GN solver, which resulted in an SD error varying from 7.74% to 50.86%.

### 2.2. Phantom experiments

After the simulations, phantom experiments were conducted to validate the proposed method. To test the robustness of the method to noisy data and to previously unseen patterns, the ANNs were trained with and without the presence of noise in the data. Electrical current was injected using two adjacent electrodes and measured with other pairs of adjacent electrodes.

The images obtained from real EIT measurements are shown in Figure 6 and Supplementary Figures 3 and 5. For the simulations, results from the proposed method were compared with those from three other methods that are widely used in EIT imaging.

The images obtained from phantom data with the linear one-step GN method (Figure 6, 1st column) showed results similar to those obtained from the simulation data. Although the conductivity distribution showed the presence of a reconstructed target, it did not clearly show its location. Instead, the reconstruction tended to show a target pushed towards the outer boundary of the FE model. If the distance between the probe and the target was large, as in Figure 6 (s), the target object might not be clearly visible. The nonlinear iterative PDIPM method also produced results similar to those obtained in the simulation study. The conductivity distributions shown in Figure 6 (2nd column) showed the presence of a reconstructed target located close to the expected location, as identified by the green ellipses on the images.

For the simulations, the reconstructions showed rough boundaries for the part of the target that was close to the probe, and smoothness on the opposite part that was close to the outer boundary. This set of results can be explained by the fact that current density injected from the probe tends to be smaller when the target is farther away from the probe, thus, its EIT sensitivity and reconstructed images tend to be better when the target is closer to the probe. Also, as the target gets farther from the probe (as shown in Figure 6 (t)), the presence of noise starts to influence the quality of the reconstruction. In that case, the resulting image shows the presence of the reconstructed target at a slightly different position. It can be assumed that with this reconstruction method, increasing the distance will give a poorer reconstruction, which may lead to wrong interpretations of the images.

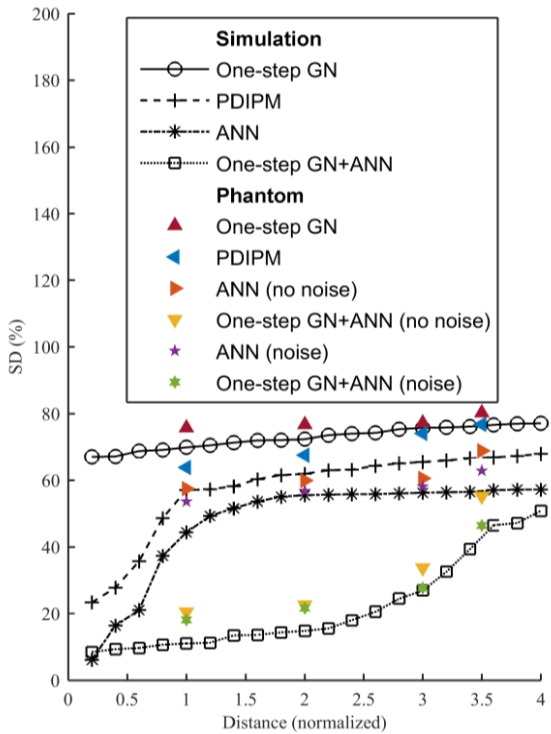

Figure 5 SD vs. distance from probe to target, as obtained from simulation and phantom experiments with four different methods. The distance is normalised based on the radius of the probe. Detailed values are given in Supplementary Table 3 (simulations) and Supplementary Table 6 (phantom).

Next, EIT measurement data were sent to an ANN that was used as an inverse solver. This ANN was trained without considering the inevitable presence of noise in the measured data. The images obtained with this ANN are shown in Figure 6 (3rd column), in which it appears that the ANN had difficulty generating an accurate approximation of the conductivity distribution. It appears that the ANN was unable to accurately estimate the size and location of the target, and instead showed large artefacts.

This result could be theoretically expected from the ANN, in that it is known to perform poorly in situations involving noise or previously unseen patterns [4]. As the one-step GN method usually offers strong robustness against noise, it was expected that the output image obtained from this method would not contain large amounts of noise, even if the measured voltages contained noise. An ANN was trained to solve the post-processing problem resulting from noise-free voltages, and this ANN was then applied to the images obtained from the phantom data (as shown in Figure 6 (1st column)). The output of this ANN is shown in Figure 6 (4th column). The reconstructed target seems generally smaller than the expected target. However, the reconstruction shows the presence of the target at the expected location, and the smoothness and artefacts are significantly reduced compared to the images generated by the PDIPM solver or the ANN used as an inverse solver.

Following the method described in the methods section, some noise, similar to the actual noise present in our measured data, could be generated artificially. Generating such noise artificially allows one to create simulation data that are more realistic to the phantom data. Then, this noise was added to the resulting voltages obtained from simulations of the forward problem. By adding noise into the measured voltages, it became possible to train the ANNs more efficiently. In fact, when adding noise, the need for extrapolation of the ANN was significantly reduced. This result is illustrated in Figure 6 (5th column), where the ANN being used as an inverse solver was trained with noisy voltages. In Figure 6 (5th column), the ANN shows the presence of a reconstructed target at the expected location, with a few distortions such as those seen in the simulation case shown in Figure 2 (3rd column). In simulations where the distances (edge-to-edge) between the probe and the target increased, the ANN had serious difficulty accurately imaging the expected target.

Following the addition of noise in an ANN inputs, the ANN was trained to be used as a post-processor. This method of using the ANN as a post-processor reduced the need for extrapolation to previously unseen patterns. The results obtained with such an ANN are shown in Figure 6 (6th column). The presence of a reconstructed target can clearly be seen at the expected location. The distortion at the boundaries of the target appears to be low, as expected. In the images obtained with the post-processing method and with an ANN trained on noise-free data, ringing artefacts are not visible. The size of the reconstructed target approximates the size of the initial target, as shown by the green ellipses in Figure 6. The proposed post-processing method obviously yielded the highest quality reconstruction.

The NADE, |ΔRES| and SD errors were then computed. For the simulations, the one-step GN method gave a high NADE, of at least 3.89. The PDIPM method generally offered higher quality and a minimal error level for targets that were physically close to the probe, with a NADE of 1.63. Although the ANN used as an inverse solver gave a minimal error of 1.91, this solution showed quickly declining quality as the distance to the target increased, with the NADE rising to around 2.5. Finally, the proposed method resulted in the lowest NADE errors. The low error levels obtained with an ANN trained by noise-free data indicated the robustness of the proposed method. The NADE errors obtained from phantom experiments are shown in Figure 3 and in Supplementary Table 4.

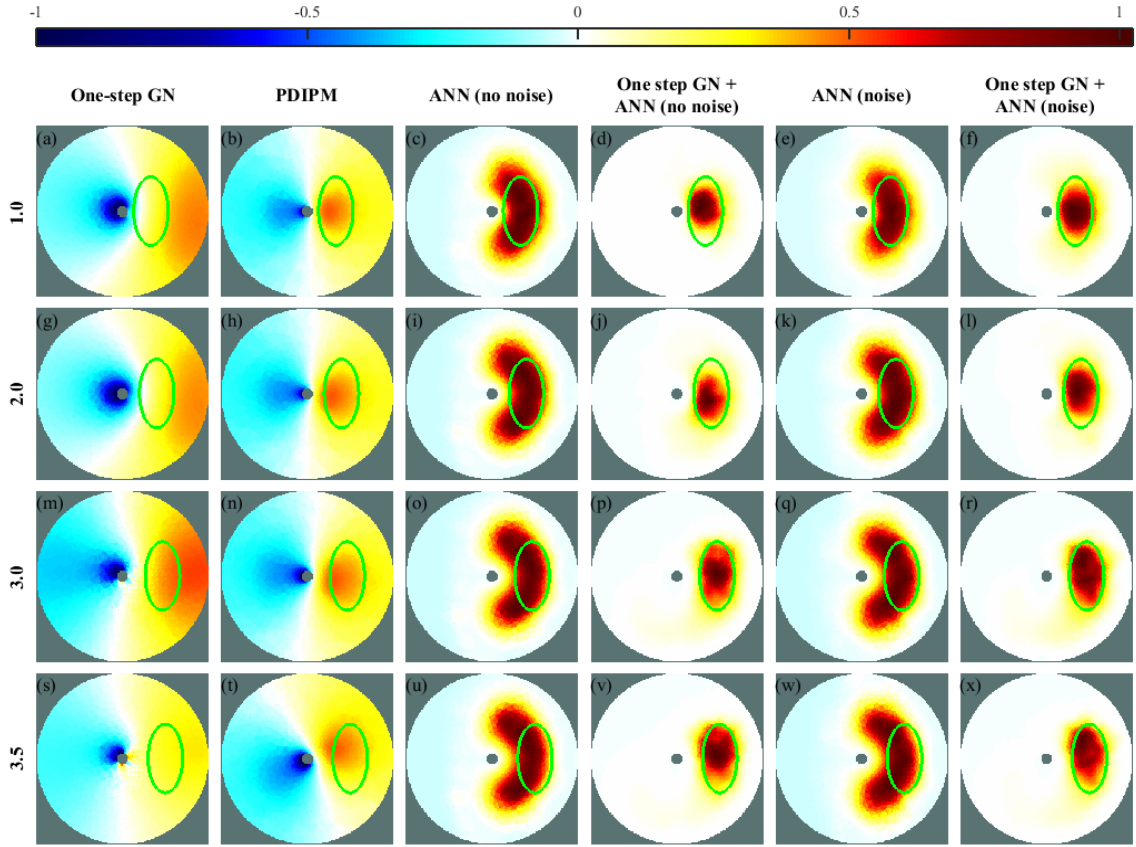

Figure 6  EIT reconstructions of the STN-shaped target built with phantom data derived using four different methods, namely the one-step GN (1st column), the PDIPM (2nd column), the ANN used as inverse solver and trained without considering the presence of noise (3rd column), the proposed post-processing as trained without considering the presence of noise (4th column), the ANN used as an inverse solver and trained by considering the presence of noise (5th column) and the proposed post-processing as trained by the presence of noise (6th column). The normalised distance between the probe and the target is shown on the left. The green ellipse represents the expected location of the STN target. The normalised resistivity distribution is shown on top. The corresponding errors are shown in Figure 3 (NADE), Figure 4 (|ΔRES|) and Figure 5 (SD). The 3D reconstructions are shown in Supplementary Figure 3 and 5. Generated with EIDORS v3.8: http://eidors3d.sourceforge.net/

Similarly, the low level of |ΔRES| errors obtained for the proposed method indicated the quality of the reconstructions obtained with this method. These |ΔRES| errors are shown in Figure 4 and Supplementary Table 5. The linear one-step GN, the PDIPM solver and the ANN used as inverse solver all gave errors of at least 19.57%, and the proposed post-processing method trained with noisy voltages gave significantly lower error levels, varying from 2.48% to 17.28%. If trained without including noise in the simulated voltages, the post-processing method gave an |ΔRES| error level up to 4% higher.

As with the NADE and the |ΔRES|, the SD errors showed a strong advantage from the proposed reconstruction method (as shown in Figure 5 and Supplementary Table 6). If trained from noisy voltages, the post-processing method always gave error levels below 47%, and the other reconstruction methods always gave error levels above 62%.

### 2.3. Brain stimulation of different regions

Although DBS surgery is usually used for placing permanent lead at the STN region for Parkinson's disease, this technology is also applicable to other regions of the brain. This section demonstrates that the proposed method can also be applied for imaging different regions. Phantom experiments were conducted using a cylindrical target twice the size of the experimental probe, and the reconstructed images are shown in Figure 7 and Supplementary Figures 4 and 6.

The one-step GN solver produced an increased degree of distortion, in size and location, so it was difficult to identify the target in Figure 7 (1st column). The reconstruction showed the presence of a reconstructed target at the outer boundary rather than at the actual location of the target (shown as a green circle). There is a large ringing effect as evidence in the blue region on the left.

In this case, the PDIPM solver was still able to generate a visual estimate of the target's location as shown in Figure 7 (2nd column). Although a region of higher resistivity was visible at the expected location, there is a significant distortion of the size of the reconstructed target. The presence of noise in the measurements generated some artefacts in the images. These findings explain why some of the reconstructions obtained with this method (such as that shown in Figure 7 (t)) indicated a slightly shifted location. In other words, the reconstruction showed a target near the expected location, but the presence of noise in the measured data could induce some errors in the target's position, and therefore this algorithm proved unreliable for use in a noisy environment.

Using an ANN as an inverse solver required training the ANN with noisy data to generate an efficient EIT reconstruction. This ANN, being highly sensitive to previously unseen patterns, could not deal effectively with such patterns. Figure 7 (3rd column) shows the images obtained with this method, and with an ANN trained without considering noise. As can be seen from the reconstructions, the ANN trained without noise produced inaccurate estimates of the size or the location of the target, and large artefacts were visible.

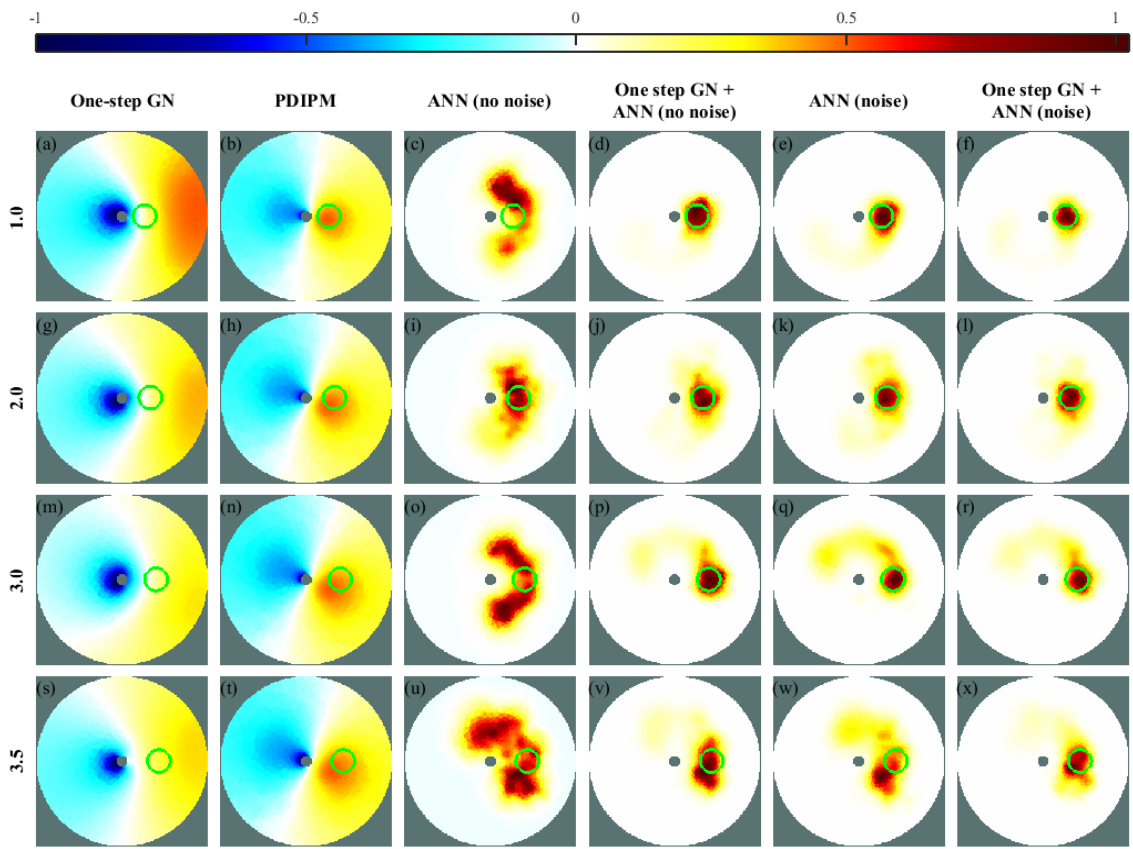

Figure 7 EIT reconstructions from phantom data generated by four different methods, namely the one-step GN (1st column), the PDIPM (2nd column), the ANN used as inverse solver and trained without considering the presence of noise (3rd column), the proposed post-processing trained without considering the presence of noise (4th column), the ANN used as an inverse solver and trained by considering the presence of noise (5th column) and the proposed post-processing as trained by the presence of noise (6th column). The normalised distance between the probe and the target is shown on the left. The green circle represents the expected location of the target. The normalised resistivity distribution is shown on top. The corresponding errors are shown in Supplementary Table 7 (NADE), Supplementary Table 8 (|ΔRES|) and Supplementary Table 9 (SD). The 3D reconstructions are shown in Supplementary Figure 4 and 6. Generated with EIDORS v3.8: http://eidors3d.sourceforge.net/

When using the proposed post-processing, the linear algorithm used at the first step of the reconstruction was expected to reduce the influence of noise. The ANN was then used as a post-processor gave an accurate solution of the inverse problem even if trained without considering the presence of noise. The EIT reconstructions obtained with this method are shown in Figure 7 (4th column), and they show that this method did not produce visible artefacts. Furthermore, the proposed method generated significantly less ringing effect than the PDIPM solver, or less artefacts and distortion than the ANN used as an inverse solver but trained without adding noise in the voltages. This set of observations validated the idea that the post-processing method was more robust than the ANN used as an inverse solver.

If the noise model is known, then it is possible to use that model to generate more realistic training data. The reconstructions obtained with an ANN used as an inverse solver (trained with noisy data) are shown in Figure 7 (5th column). As can be seen from the images, this method could generate efficient reconstructions. As the target moved farther from the probe, the signal became weaker, and the SNR grew smaller. That finding explained why this solution generated the artefacts shown in Figure 7 (q and w). The PDIPM solver, however, tended to produce a reconstructed image with a poor estimate for the location of the inhomogeneity.

Finally, Figure 7 (6th column) shows the EIT reconstruction of phantom data using the post-processing method and an ANN trained with reconstructions from noisy data. In that case, the proposed post-processing method gave the highest-quality. The target inserted into the phantom appeared to be at the expected location, as represented by the green circles in the images. Compared with other reconstruction methods, it can be seen that this proposed method generated no large artefacts, even as the distance from the probe increased.

The NADE, $|\Delta RES|$ and SD errors were calculated. The NADEs are shown in Supplementary Table 7. As for the STN region, the proposed method significantly reduced the artefacts in the images, which in turn reduced the levels of error. Using the proposed method, an ANN trained with noisy data gave the lowest NADE (from 0.47 to 1.10), and an ANN used as an inverse solver gave error levels varying from 0.58 to 1.45. If trained without noise, the ANN used as an inverse solver gave a degenerated solution, with an error higher than the error level for the one-step GN solver. However, the proposed method could still give a high-quality reconstruction, even when the ANN had been trained with noise-free data. In that case, the NADE increased from 0.50 to 1.47.

Supplementary Table 8 shows the $|\Delta RES|$ errors. As in the previously reported tests, the results showed the superior stability of the proposed post-processing method. The lowest errors were obtained from the proposed method and an ANN trained with the presence of noise. The SD errors, shown in Supplementary Table 9, confirmed these observations.

## 2.4. Computation resources

In this section, the reconstruction times and the memory required to solve the inverse problem with the different methods are compared. The CPU time and memory are given in Table 1.

Due to the large size of the mesh (above 11000 nodes), the reconstruction matrix used to solve the inverse problem with the one-step GN required more than 160 GB and more than eleven hours to compute. However, this step could be carried out before collecting the data. In that case, the reconstruction consisted only of a matrix product, which was solved in 319 milliseconds with less than one GB of memory. The iterative PDIPM solver required large computation capacity for each iteration, which made the process extremely time consuming. In 3D applications, this method required more than two hours and 120 GB of memory for converging to a solution. However, an ANN used as an inverse solver could solve the problem in less than two seconds with 1.3 GB of RAM.

Table 1 CPU time and required memory to solve the 3D inverse problem by using four different methods

|             | One-step GN | PDIPM    | ANN   | One-step GN + ANN |
|-------------|-------------|----------|-------|-------------------|
| Time (s)    | 0.319       | 7332.656 | 1.82  | 2.918             |
| Memory (GB) | 0.858       | 120.243  | 1.262 | 3.606             |

Finally, the proposed post-processing method combined the one-step GN with an ANN, which solved the inverse problem in less than three seconds and required less than 4 GB of memory. This amount of memory is available on any modern computer. studies.

## 3. Discussion

This paper proposes a novel reconstruction method for a 3D open-domain EIT. In such a configuration, the iterative solutions are extremely slow, and they cannot be applied to real-time imaging. Linear solutions can be applied in real-time, but large distortions may result, and the visual quality of those methods tends to be poor.

According to the test results, the proposed post-processing method resulted in high-quality image reconstruction in near real-time by taking advantage of the higher dimension of the input data. Phantom experiments showed that using an ANN as an inverse solver usually offered weak robustness to noise. However, the proposed method offered a very high robustness to noise present in the phantom even if the ANN was trained without any prior exposure to noise in the model. This method showed considerable advantages for the proposed method for enabling future applications of EIT during DBS surgery, in which the noise model remains mostly unknown.

This paper introduces a novel approach to solve the EIT inverse problem. Simulation studies and phantom experiments showed the high efficiency of the proposed method, as compared to three other reconstruction methods that are widely used in EIT. Reconstructions were obtained within three seconds, which is an acceptable time for such an application, as the DBS microelectrodes move relatively slowly during DBS surgery. Although brain surgery is an active research area, little previous work has been done to model the electrical properties of the various brain tissues, and therefore at the present time, simulations are likely to contain modelling errors. Although the modelisation of the problem was not perfect, it remained possible to apply the proposed post-processing method to get high-quality imaging. The regularisation effects of the linear solver is clear from the experiments. Future applications will focus on more complex modelling of the brain, and on conducting further experiments with more realistic phantoms or with animal studies.

## 4. Method

### *4.1. Simulations and the inverse problem*

In EIT simulations, the problem can be divided into two different problems, namely the forward and the inverse problems [17]. The forward problem starts from a given conductivity distribution and a known current injection pattern, and the solution gives the corresponding voltages at the electrodes.

The goal of the inverse problem is to determine the properties of the object present in a volume conductor, as based on measurements made at the boundary of the domain [18]. This reconstruction problem is the most challenging to solve in EIT because a small change in electrical conductivity of the medium can result in a large change in the electrical potential measured at the boundaries [19].

Several linear and nonlinear solutions have been proposed to solve this inverse problem. Each solution has both advantages and drawbacks. Linear algorithms, which are usually used in biomedical EIT applications such as the one-step GN [20], usually have strong robustness against noise, movement or modelling error, and are regarded as reliable for biomedical use. However, their underlying mathematical model assumes a linear variation from an initial estimate.

Nonlinear iterative methods, such as the PDIPM, are theoretically able to give a nonlinear estimate of the conductivity distribution. Although those methods are more efficient, they are also known to be more sensitive to noise and movement artefacts. In addition, iterative methods are extremely time-consuming and are therefore not suitable for real-time applications.

Algorithms based on an ANN are able to generate nonlinear solutions to any kind of nonlinear problem, after a training process in which the training algorithm automatically finds the best set of weights and biases for the specific problem at hand. This method usually gives satisfactory results, which are close to the original distribution under noise-free conditions [21]. However, using AI to solve an EIT inverse problem may also give non-satisfactory results, especially when the ANN has to deal with unseen input data, or if it cannot correctly evaluate a specific set of measured voltages. In practical applications, this problem can result from a poorly trained ANN, or an ANN trained with the wrong data. A common mistake is to simplify the generation of training data from simulations, without considering the presence of noise in the data measured. It then becomes necessary to determine a correct way to model the noise, which strongly depends on the hardware used.

In this paper, a novel combination of linear approximations and nonlinear optimisations for a three-dimensional (3D) EIT were proposed and tested. After solving the inverse problem using a linear algorithm such as one step GN, an ANN was used to enhance the quality of an EIT image. The ANN aimed to reduce both the artefacts and the presence of smoothness generated by the linear inverse solver. Applying the ANN on the conductivity distribution in the mesh (in other words, after solving the inverse problem with a linear inverse solver), allowed greater convergence and minimised the risk of failure. In fact, the ANNs used in a nonlinear reconstruction method were very sensitive to noise and to other artefacts present in the data. Linear algorithms are known to be resistant to noise, and are expected to eliminate the artefacts generated by noise or modelling errors. After application of such algorithms, it was possible to train an ANN without including noise or distortion in the training data, and still obtain a reduced risk that the ANN would detect an inaccurate pattern. This process, therefore, reduced the need for good extrapolation [22]. The proposed method thus decreased the modelling effort necessary to train the ANN, compared to an ANN that directly processes the measured voltages.

### *4.2. Finite element method*

A large 3D finite element (FE) model was created with a probe at its centre. The probe consisted of four layers of eight electrodes each, with each electrode having a 5 mm * 5 mm aluminium sheet. The prototype probe used for the experiments had a radius of 2.5 cm. The size of the model used to solve the problems numerically was normalised to 1, and the size of the STN was varied accordingly. Currently, the most common DBS systems use a thin circular probe with a radius of 1 mm or less.

Previous work in estimating the shape and size of the STN region had determined that the region was elliptical. Although its real size could vary among different patients, a typical STN region for an elderly patient could be roughly represented by an ellipsoidal shape with radii of 4, 6 and 9 mm. In the simulations and in the phantom experiments, the ratio of the probe to target size was similar to the ratio of the probe to STN region size that is likely to be found in future clinical studies. Maximal imaging distances strongly depended on the size of the probe, and distances were therefore given as ratios of the radius of the probe [23].

When solving the inverse problem, reverberation effects may reduce the quality of the reconstructed EIT image [24]. This phenomenon results from the finite size of the FE model and the presence of an outer boundary. To avoid any reverberation effects, the FE model used was significantly larger than the maximum distance between the probe and the target object. Although the centre of the object was no more than ten times the radius of the probe, the perimeter of the open-domain EIT tank was 70 times larger than the probe. To speed up the computation of the inverse problem, a fine non-uniform-sized mesh was applied around the probe and the ROI.

To train the ANN, a training data set of 3000 images was generated, using an elliptical object similar to that of the STN region. This object was placed in the tank close to the probe. The target objects could be located anywhere within a radius of up to ten times the radius of the probe. Images were obtained by using the FE method, with a large mesh of 14014 nodes and 66482 elements. That mesh was refined around the target object and around the DBS probe to allow higher precision in that area. The FE model used for the reconstruction is shown in Supplementary Figure 1. Reconstruction algorithms, EIDORS [25] and MATLAB's ANN toolbox were used. The CPU times were measured on an Intel Core i7 6700K CPU aided by 64 GB of RAM and Ubuntu Linux.

### 4.3. Simulation settings

For simulations, some efforts have been made on re-creating a more accurate scenario in terms of electrical conductivities. For example, the STN target was not a pure insulator but has the electrical conductivity of the actual STN target, as measured in [26]. To summarize, authors found that the STN target (gray matter) has different electrical conductivity than the surrounding white matter. The electrical conductivity of the STN region was set to 0.3S/m, surrounded by white matter having an electrical conductivity of 0.15S/m. Those values allow for better modelling of the brain electrical properties and in turn in a better simulation of the actual propagation of the electrical current in the brain area. An electrical current of 5 microampere at the frequency of 1kHz was utilized in those simulations. The configuration was tested at different intensities without noticeable changes in simulations.

### 4.4. Data acquisition

Phantom data were acquired by using the data acquisition system described in [27]. This system used pairs of two adjacent electrodes (which were located on the same layer) for current injection and voltage measurements. For each current injection, 32 different voltages were measured by using the 32 electrodes at the boundaries of the probe. The current source was then moved to the next pair of adjacent electrodes, and 32 further measurements were made. A total of 1024 measurements were used for image reconstruction. Although this method did not give 1024 independent measurements, the common practice was followed by keeping all the measurements to reduce the influence of noise [28]. Those measurements obtained with the electrodes used for current injection were not kept for the reconstruction. The probe was inserted into a large water tank to avoid reverberation effects and to minimise the influence of the physical boundary of the phantom. A photo of the water tank containing the experimental probe is shown in Supplementary Figure 2. An electrical current of 5 mA at 1kHz was applied using two adjacent electrodes made of aluminium. Each electrode consists of a 5mm*5mm piece of aluminium, which gives a current density of 200A/$m^2$. While the electrodes on an actual DBS probe will be about fifty times smaller than our model, they are capable of injecting and measuring an electrical current with a precision below the microampere [29]. Previous work from Hannan and colleagues showed that injecting AC current into the brain at 1.7 kHz up to a current density of 354 A/$m^2$ is safe and does not cause tissue damage [30]. Therefore, the state of the art of the hardware for DBS can safely be utilized for this application [31].

### 4.4.1. Noise estimates

The presence of noise and errors resulting from inaccurate modelling were taken into consideration during the phantom experiments. Although most of the noise was cancelled by using a tenth-order bandpass filter (which was centred on the frequency of the injected current), a certain amount of noise was still present in the acquired data. To reduce the need for the ANN to extrapolate from unknown data due to the presence of noise, it was advantageous to train the ANN with noisy data, similar to the data acquired from the phantom. Although the proposed method makes the network more robust against noise, it still benefits from more realistic training data.

The amount of noise was not fixed, and it strongly depended on both the physical distance between the current source and the electrodes used for measurement. Noise was estimated independently for each measurement. Each current injection consisted of a sinusoidal waveform at a frequency of 2 kHz. Noise was estimated by comparing the measured data to a simulated sine wave. By estimating the amounts of noise in the measurements before filtering, and by assuming the noise to be a white Gaussian noise, it was possible to estimate the noise levels for the different measurements. This procedure is described in greater detail in [16]. To summarize, it was found that for measurements using a pair of two adjacent electrodes located nearby the two adjacent electrodes acting as a current source, the SNR could be above 50dB. However, when the voltages were measured using adjacent electrodes located on the other side of the DBS probe, the SNR could be lower than 10dB. In short, the further the current source away from the measuring electrodes, the lower the SNR. A complete description of the measured noise is out of the scope of this paper, and is described extensively in [32].

*4.5. Artificial Neural Networks*

ANNs are a series of algorithms that have been proven extremely powerful to solve a nonlinear problem, including inverse problems within a FE model. Given sample data, the model is capable of learning the expected output given a specific input data [33]. In our case, given the measured voltages (or the reconstruction from the linear model in the case of the post processing solution), the ANN is learning to reproduce the original conductivity distribution. If training is learning the correct problem, then the resulting ANN is capable of reproducing the original distribution within the FE model. This training step is being done with the backpropagation algorithm, which adjusts the weights of the model after iterating through a batch a simulated training data. After processing a batch of training data, the output of the ANN is compared to the expected output (the original distribution at each node of the FE model) and the mean square error is calculated. From this error, the backpropagation algorithm is capable of optimizing the weights of the model [34]. A decreasing error means that the model is actually learning to solve the problem. In order to ensure that the results are not biased, the same procedure and a similar model was utilized for both ANN as an inverse solver and ANN as a post processing. The only difference being the input data. After processing all the training samples, we start again until the convergence (during validation) stops decreasing. A common problem with training ANNs is that the model overfits on the training data but does not generalize well on new patterns unseen during the training phase, even if they are very close to the previously seen patterns. To avoid this problem, the dataset is generally split into two, called training dataset and validation dataset. After processing the whole training dataset, the validation dataset is fed to the ANN, and the resulting mean square error is expected to be lower than the error at the previous iteration. In this application, ten percent of the whole dataset was utilized for validation. All of the models used in this paper (as an inverse solver or as a post-processor) are made of three layers and five thousand hidden neurons. The model loss (mean square error) was decreasing, indicating that the model was converging to the expected result. Radial basis functions (RBFs) were used in the hidden layer of neurons. The output layer was made of a linear transfer function. Research has shown that this configuration of an ANN is capable of high quality EIT reconstruction from biomedical data. For an ANN used as an inverse solver, the input layer has a number of neurons equivalent to the number of measurements (928 neurons). For the post-processing application, the input layer has a number of neurons equivalent to the number of nodes present in the FE model (or 14014 in this application). In this study, in both cases, the hidden layer was made of 2,000 neurons with a RBF transfer function. Finally, the output layer outputted the estimate of the conductivity distribution within the FE model and contained a number of neurons equivalent to the number of nodes present in this model.

For training neural networks, electrical targets having the conductivity of the STN region surrounded by white matter were generated. For each target, the location, orientation, and direction of that region were chosen randomly with a maximal distance probe – target (edge to edge) of ten times the diameter of the probe. More than 2000 conductivity distributions were generated that way. From those conductivity distributions, and given a known current injection scheme, the forward problem could be solved. This forward problem simulated the electrical voltages measured at the electrodes, to which an additional white Gaussian noise can be added in order to produce a more realistic data. Given those measured voltages, one can approximate the conductivity distribution by solving the inverse problem. At this stage, we could train the ANN used as a direct inverse solver by submitting the simulated voltages as input and the corresponding conductivity distributions as output. Then, given the simulated voltages, the inverse problem was solved using the one step GN method. That step gave the input data of the ANN used as a post processing technique. For this model, the input of the neural network is the solution to the inverse problem, while the output is the original conductivity distribution. Finally, in order to avoid committing the inverse crime, different FE models were utilized to solve the forward and inverse problems.

*4.6. Errors definitions*

To obtain an objective definition of image quality, an objective error function has to be defined. Several methods have been proposed to define such a function for medical imaging applications [35]. Among these definitions, the |ΔRES| and the SD are particularly relevant to the proposed application [36]. In the original definition, the resulting conductivity distribution in the mesh was thresholded at 25% of the maximum conductivity in order to create a binary image. Since this work is using a 3D mesh, we obtained a voxel spaces and computed those errors on this voxel space instead of the image space [4]. This configuration allows for processing the error over the 3 dimensions of the reconstruction.

The NADE was also used in this work [37]. As with the |ΔRES| and SD, the NADE was determined from the thresholded 3D image and a region of interest that was twice as large as the actual target that was to be reconstructed. An additional definition, called Average Distance Error (ADE), is proposed for open-domain EIT. It measures the quality of the location of the reconstructed target. This error considers a specific ROI around the target location. This ROI has a similar shape than the target and was thresholded similarly to the GREIT errors definitions mentioned above. The ADE then compares the ROI of the binary image obtained from reconstruction after applying the thresholding. After thresholding, parts of the reconstructed image containing an object where the initial image (the simulated target) does not, and vice versa, are considered as error. Then, the area of the error

region is computed and divided by the perimeter of the elliptical target, and normalized to the diameter of the probe [23].

For all reconstructions derived either from simulation or phantom experiments, the resulting errors were compared based on the edge-to-edge distances between the probes and the target objects.

## 5. Acknowledgement

The author wishes to thank Chii-Chew Hong for designing the hardware system the EIDORS project for providing the EIT software libraries, and the Ministry of Science and Technology (MOST) of Taiwan for funding this research under grant 107-2221-E-009 -088-MY3.

Supplementary Table 1 NADE obtained from simulations for a target similar to the STN region. Some of the corresponding images are visible in Figure 2 in the original paper

| Distance (normalized) | Method | | | |
|---|---|---|---|---|
| | One-step GN | PDIPM | ANN | One-step GN + ANN |
| 0,2 | 2,69 | 1,01 | 0,54 | 0,31 |
| 0,4 | 3,02 | 1,03 | 0,57 | 0,34 |
| 0,6 | 3,13 | 1,05 | 0,67 | 0,36 |
| 0,8 | 3,44 | 1,24 | 1,10 | 0,37 |
| 1,0 | 3,51 | 1,32 | 1,28 | 0,39 |
| 1,2 | 3,76 | 1,39 | 1,56 | 0,43 |
| 1,4 | 3,78 | 1,43 | 1,86 | 0,46 |
| 1,6 | 3,88 | 1,64 | 1,95 | 0,47 |
| 1,8 | 3,93 | 1,81 | 2,02 | 0,48 |
| 2,0 | 3,95 | 2,04 | 2,04 | 0,52 |
| 2,2 | 3,97 | 2,08 | 2,06 | 0,54 |
| 2,4 | 4,06 | 2,35 | 2,06 | 0,56 |
| 2,6 | 4,13 | 2,51 | 2,07 | 0,59 |
| 2,8 | 4,17 | 2,78 | 2,08 | 0,61 |
| 3,0 | 4,20 | 2,93 | 2,10 | 0,65 |
| 3,2 | 4,23 | 3,07 | 2,19 | 0,78 |
| 3,4 | 4,27 | 3,51 | 2,20 | 1,09 |
| 3,6 | 4,32 | 3,72 | 2,22 | 1,37 |
| 3,8 | 4,36 | 3,76 | 2,28 | 1,49 |
| 4,0 | 4,38 | 4,01 | 2,30 | 1,69 |

Supplementary Table 2 |ΔRES| (%) obtained from simulations for a target similar to the STN region. Some of the corresponding images are visible in Figure 2 in the original paper

| Distance (normalized) | Method | | | |
|---|---|---|---|---|
| | One-step GN | PDIPM | ANN | One-step GN + ANN |
| 0,2 | 18,41% | 0,71% | 1,06% | 0,62% |
| 0,4 | 20,17% | 1,99% | 3,25% | 0,62% |
| 0,6 | 21,66% | 4,34% | 4,76% | 0,75% |
| 0,8 | 23,78% | 12,58% | 10,89% | 0,98% |
| 1,0 | 24,81% | 18,83% | 14,07% | 1,45% |
| 1,2 | 25,59% | 20,83% | 17,33% | 1,61% |
| 1,4 | 26,46% | 22,86% | 19,36% | 2,13% |
| 1,6 | 27,20% | 25,17% | 20,63% | 2,22% |
| 1,8 | 27,88% | 26,53% | 20,75% | 2,95% |
| 2,0 | 28,40% | 26,89% | 21,60% | 3,10% |
| 2,2 | 28,75% | 28,17% | 21,93% | 3,62% |
| 2,4 | 29,02% | 28,67% | 22,18% | 4,55% |
| 2,6 | 29,73% | 29,73% | 22,35% | 5,52% |
| 2,8 | 30,22% | 30,55% | 22,42% | 7,05% |
| 3,0 | 30,26% | 31,35% | 22,55% | 7,61% |
| 3,2 | 30,47% | 31,77% | 22,62% | 9,98% |
| 3,4 | 30,74% | 33,30% | 22,73% | 13,26% |
| 3,6 | 30,84% | 33,64% | 22,91% | 17,09% |
| 3,8 | 31,20% | 34,14% | 23,40% | 17,64% |
| 4,0 | 31,40% | 34,27% | 23,68% | 19,63% |

Supplementary Table 3 SD errors (%) obtained from simulations for a target similar to the STN region. Some of the corresponding images are visible in Figure 2 in the original paper

| Distance (normalized) | Method | | | |
|---|---|---|---|---|
| | One-step GN | PDIPM | ANN | One-step GN + ANN |
| 0,2 | 67,00% | 23,31% | 6,12% | 8,48% |
| 0,4 | 67,14% | 27,78% | 16,39% | 9,33% |
| 0,6 | 68,75% | 35,68% | 21,05% | 9,68% |
| 0,8 | 69,09% | 48,64% | 37,32% | 10,67% |
| 1,0 | 69,86% | 57,04% | 44,34% | 11,04% |
| 1,2 | 70,46% | 57,29% | 49,26% | 11,25% |
| 1,4 | 71,28% | 58,21% | 51,61% | 13,45% |
| 1,6 | 71,95% | 60,31% | 53,65% | 13,61% |
| 1,8 | 72,04% | 61,56% | 55,05% | 14,36% |
| 2,0 | 72,31% | 61,93% | 55,53% | 14,81% |
| 2,2 | 73,53% | 62,99% | 55,70% | 15,54% |
| 2,4 | 73,99% | 63,19% | 55,80% | 18,05% |
| 2,6 | 74,21% | 64,36% | 55,86% | 20,55% |
| 2,8 | 75,33% | 65,07% | 56,04% | 24,47% |
| 3,0 | 75,74% | 65,50% | 56,25% | 26,92% |
| 3,2 | 75,83% | 65,95% | 56,40% | 32,56% |
| 3,4 | 76,11% | 66,60% | 56,50% | 39,37% |
| 3,6 | 76,62% | 66,85% | 57,04% | 46,53% |
| 3,8 | 76,96% | 67,29% | 57,18% | 47,18% |
| 4,0 | 77,08% | 67,91% | 57,25% | 50,86% |

Supplementary Table 4 NADE obtained for phantom experiments for a target similar to the STN region. Corresponding images are visible in Figure 6

| Method \ Distance | One-step GN | PDIPM | ANN (training: no noise) | One-step GN + ANN (training: no noise) | ANN (training: noise) | One-step GN + ANN (training: noise) |
|---|---|---|---|---|---|---|
| 1.0 | 3.89 | 1.63 | 2.34 | 0.67 | 1.91 | 0.49 |
| 2.0 | 4.08 | 2.29 | 2.84 | 1.10 | 2.77 | 0.73 |
| 3.0 | 4.51 | 3.11 | 2.95 | 1.21 | 2.79 | 0.94 |
| 3.5 | 4.70 | 4.85 | 3.07 | 1.46 | 2.82 | 1.31 |

Supplementary Table 5 |ΔRES| obtained for phantom experiments for a target similar to the STN region. Corresponding images are visible in Figure 6

| Method \ Distance | One-step GN | PDIPM | ANN (training: no noise) | One-step GN + ANN (training: no noise) | ANN (training: noise) | One-step GN + ANN (training: noise) |
|---|---|---|---|---|---|---|
| 1.0 | 24.96% | 23.08% | 22.24% | 4.50% | 19.57% | 2.48% |
| 2.0 | 29.72% | 28.01% | 23.75% | 5.40% | 23.67% | 3.80% |
| 3.0 | 32.94% | 34.54% | 26.13% | 11.47% | 24.55% | 8.80% |
| 3.5 | 36.28% | 36.71% | 27.13% | 20.55% | 25.39% | 17.28% |

Supplementary Table 6 SD obtained for phantom experiments for a target similar to the STN region. Corresponding images are visible in Figure 6

| Method \ Distance | One-step GN | PDIPM | ANN (training: no noise) | One-step GN + ANN (training: no noise) | ANN (training: noise) | One-step GN + ANN (training: noise) |
|---|---|---|---|---|---|---|
| 1.0 | 75.68% | 63.82% | 57.43% | 20.50% | 53.64% | 18.13% |
| 2.0 | 76.68% | 67.55% | 59.93% | 22.59% | 56.83% | 21.67% |
| 3.0 | 77.16% | 74.06% | 60.56% | 33.71% | 58.11% | 27.68% |
| 3.5 | 80.17% | 76.81% | 68.85% | 55.22% | 62.88% | 46.40% |

Supplementary Table 7 NADE obtained for phantom experiments for a small target. Corresponding images are visible in Figure 7

| Method \ Distance | One-step GN | PDIPM | ANN (training: no noise) | One-step GN + ANN (training: no noise) | ANN (training: noise) | One-step GN + ANN (training: noise) |
|---|---|---|---|---|---|---|
| 1.0 | 1.75 | 1.77 | 1.79 | 0.50 | 0.58 | 0.47 |
| 2.0 | 1.79 | 1.88 | 1.82 | 0.81 | 1.01 | 0.89 |
| 3.0 | 1.85 | 2.23 | 1.86 | 1.15 | 1.31 | 0.95 |
| 3.5 | 1.93 | 2.59 | 2.04 | 1.47 | 1.45 | 1.10 |

Supplementary Table 8 |ΔRES| obtained for phantom experiments for a small target. Corresponding images are visible in Figure 7

| Method \ Distance | One-step GN | PDIPM | ANN (training: no noise) | One-step GN + ANN (training: no noise) | ANN (training: noise) | One-step GN + ANN (training: noise) |
|---|---|---|---|---|---|---|
| 1.0 | 41.27% | 34.34% | 29.92% | 10.89% | 9.74% | 8.73% |
| 2.0 | 45.69% | 46.64% | 38.38% | 14.86% | 13.75% | 10.48% |
| 3.0 | 50.27% | 52.77% | 48.44% | 18.34% | 18.65% | 14.08% |
| 3.5 | 52.67% | 56.11% | 58.81% | 23.00% | 29.21% | 15.46% |

Supplementary Table 9 SD obtained for phantom experiments for a small target. Corresponding images are visible in Figure 7

| Method \ Distance | One-step GN | PDIPM | ANN (training: no noise) | One-step GN + ANN (training: no noise) | ANN (training: noise) | One-step GN + ANN (training: noise) |
|---|---|---|---|---|---|---|
| 1.0 | 76.94% | 72.52% | 63.79% | 46.19% | 44.44% | 35.95% |
| 2.0 | 77.16% | 77.68% | 64.70% | 48.70% | 49.24% | 39.54% |
| 3.0 | 78.02% | 78.85% | 70.16% | 50.18% | 60.71% | 45.83% |
| 3.5 | 79.31% | 80.94% | 75.53% | 56.30% | 70.93% | 49.52% |

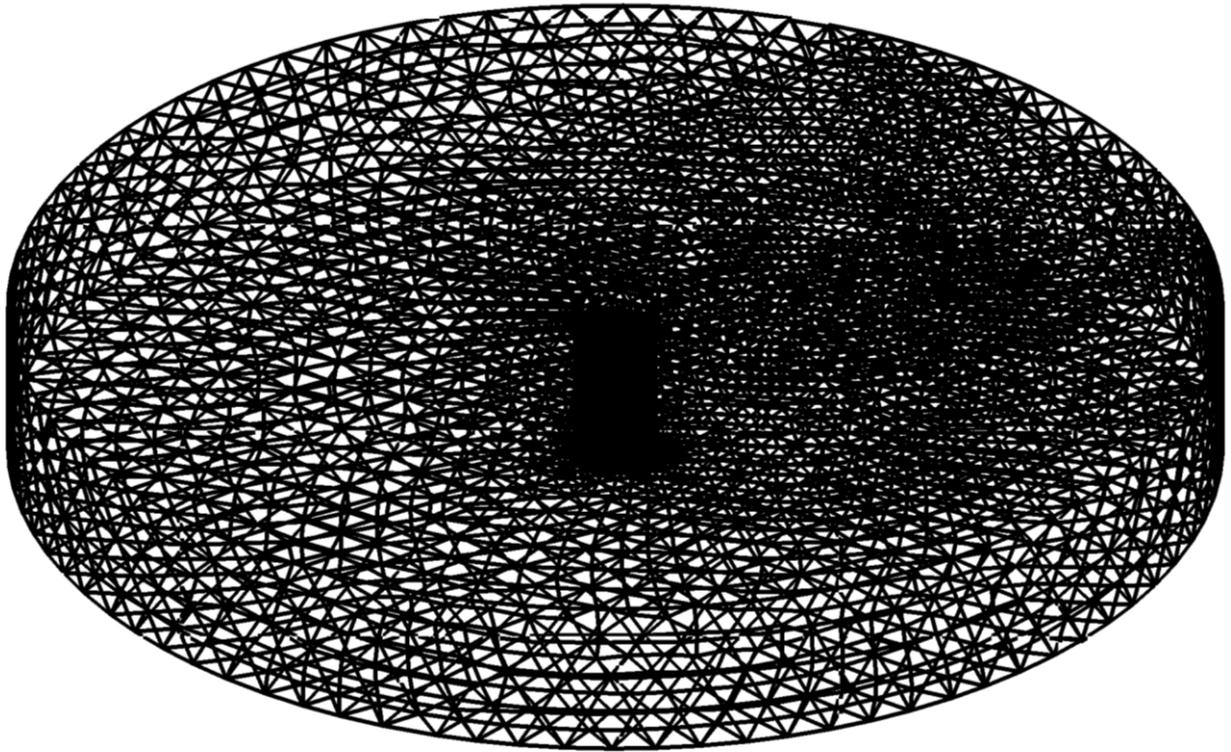

Supplementary Figure 1 The FE model used to solve the open domain 3D EIT problem. The probe, with 4 layers of 8 electrodes each, is located at the middle of the model

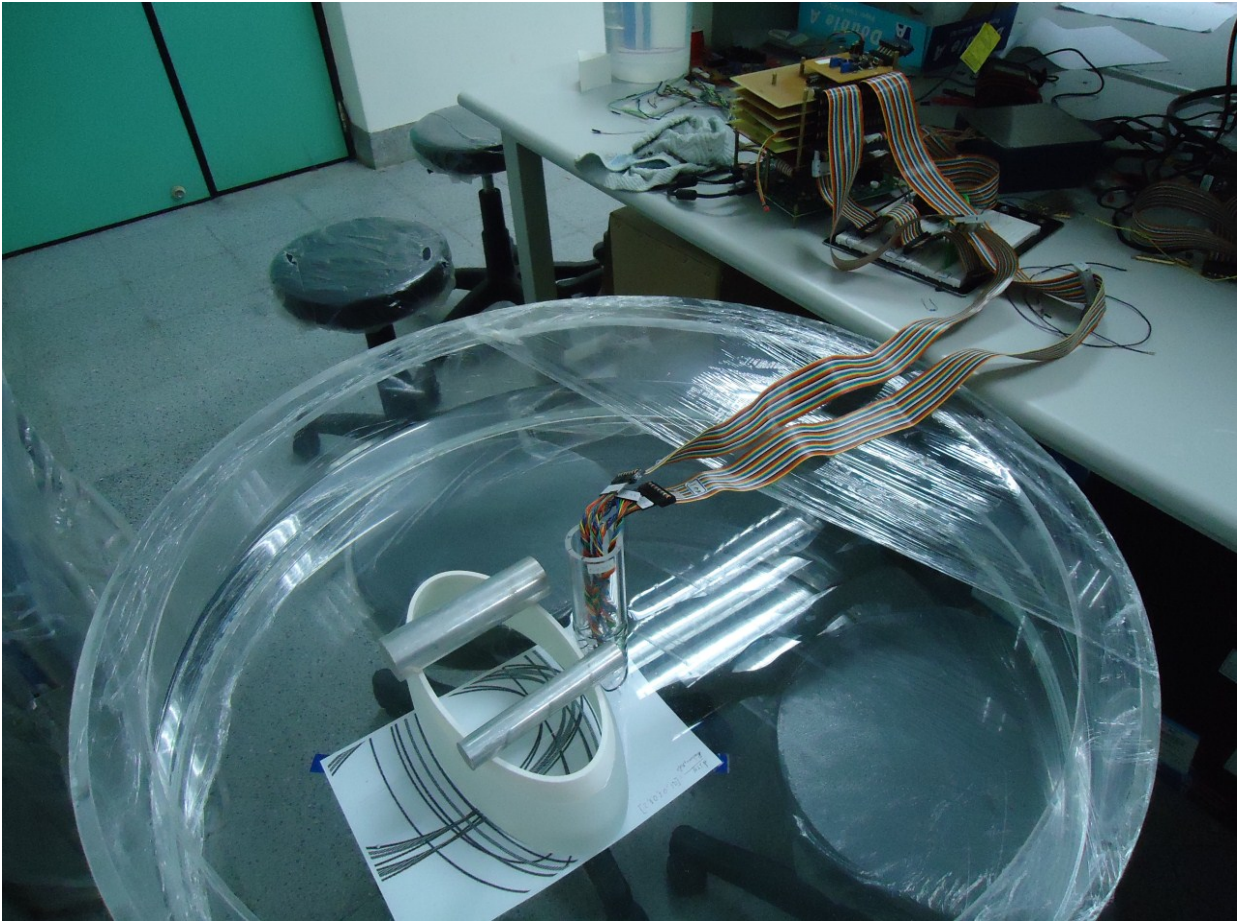

Supplementary Figure 2 The experimental probe submerged in ionized water in the middle of the water tank

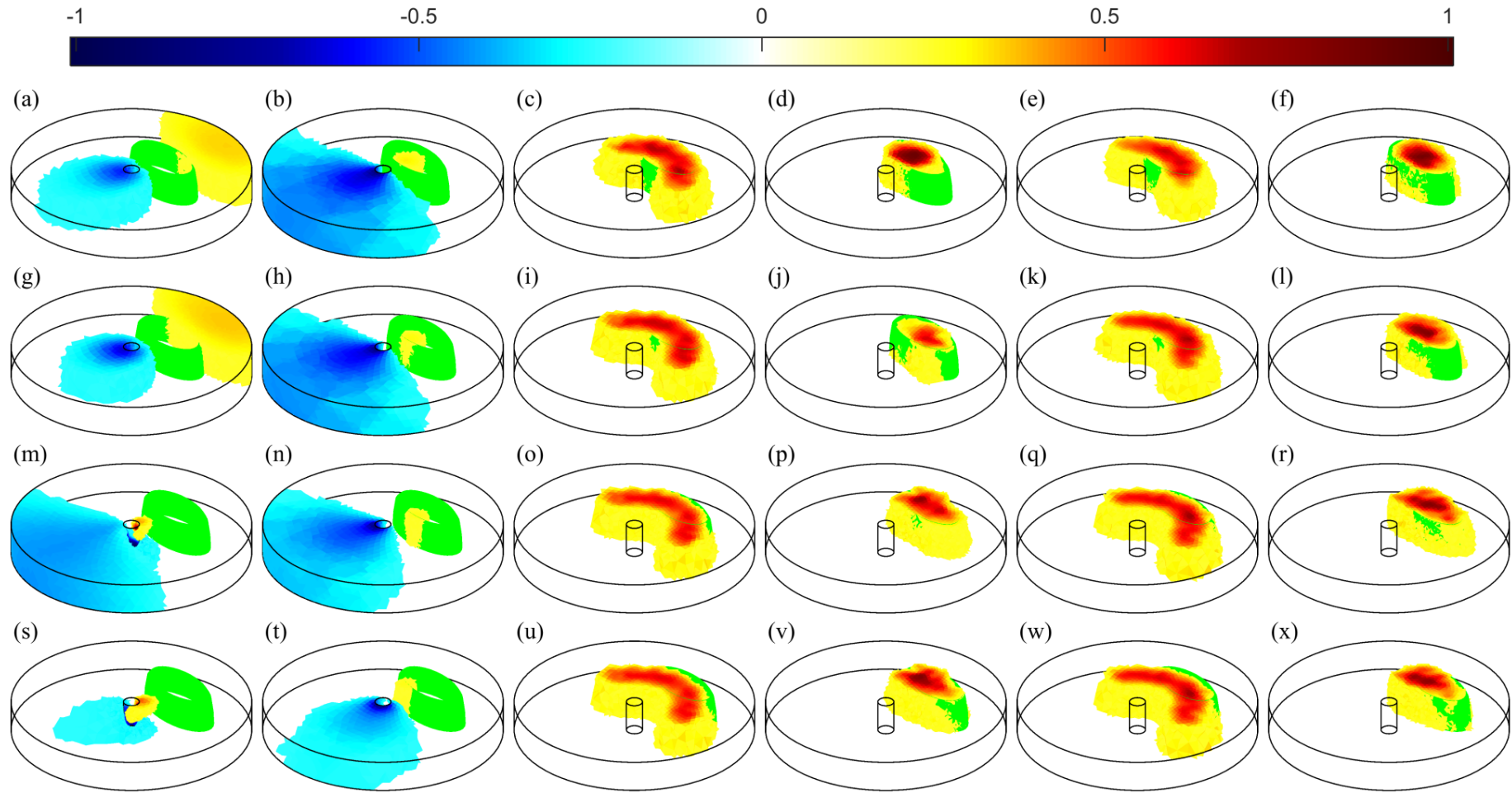

Supplementary Figure 3 EIT reconstructions of the STN-shaped target built with phantom data derived using four different methods, namely the one-step GN (1st column), the PDIPM (2nd column), the ANN used as inverse solver and trained without considering the presence of noise (3rd column), the proposed post-processing as trained without considering the presence of noise (4th column), the ANN used as an inverse solver and trained by considering the presence of noise (5th column) and the proposed post-processing as trained by the presence of noise (6th column). The normalised distance between the probe and the target are, from top to bottom: 1.0, 2.0, 3.0, and 3.5, respectively. The green ellipse represents the expected location of the STN target. The normalised resistivity distribution is shown on top. The finite element model is not shown for better visibility.

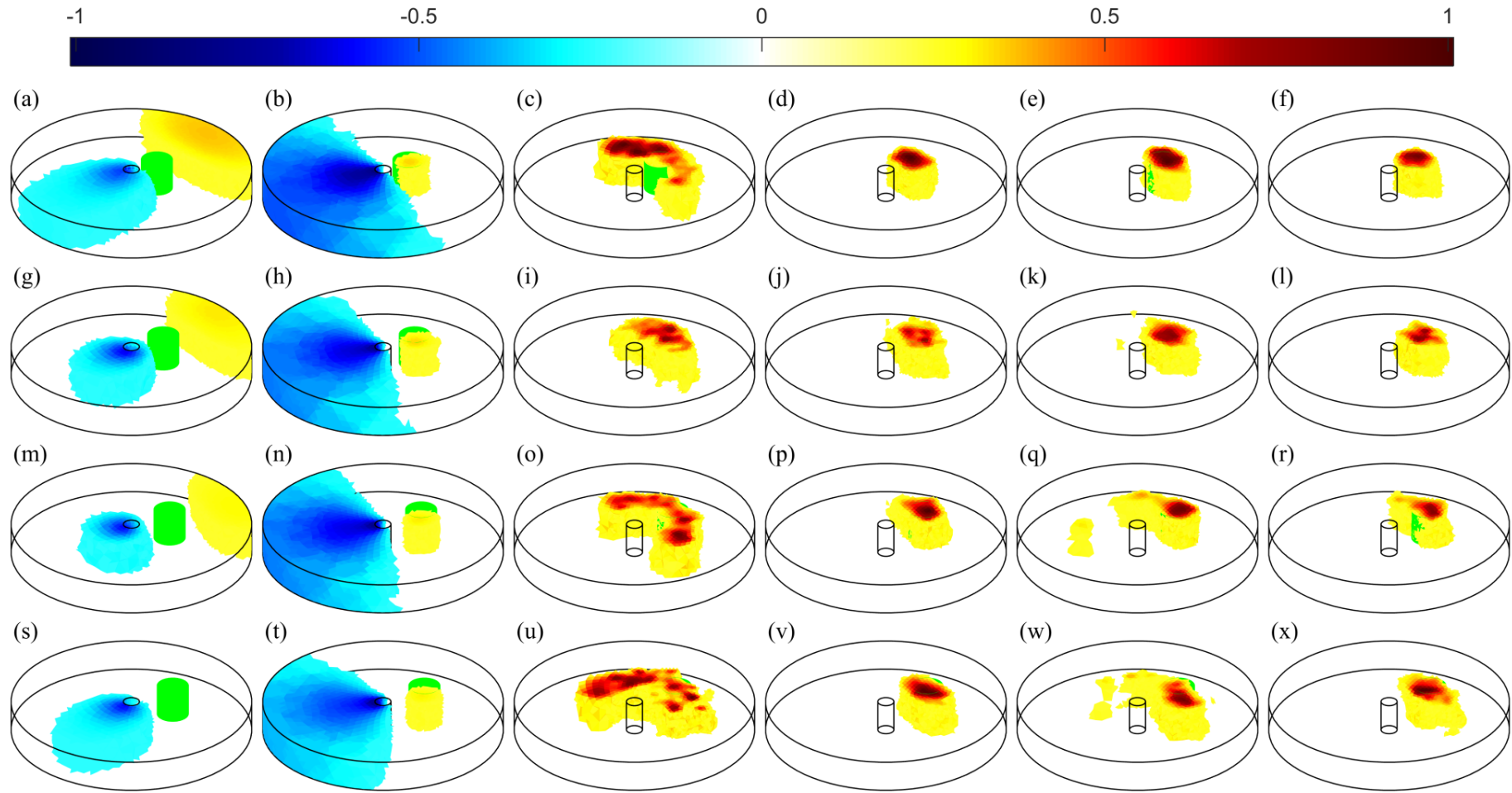

Supplementary Figure 4 EIT reconstructions of a cylindrical target from phantom data generated by four different methods, namely the one-step GN (1st column), the PDIPM (2nd column), the ANN used as inverse solver and trained without considering the presence of noise (3rd column), the proposed post-processing trained without considering the presence of noise (4th column), the ANN used as an inverse solver and trained by considering the presence of noise (5th column) and the proposed post-processing as trained by the presence of noise (6th column). The normalised distance between the probe and the target are, from top to bottom: 1.0, 2.0, 3.0, and 3.5, respectively. The green ellipse represents the expected location of the STN target. The normalised resistivity distribution is shown on top. The finite element model is not shown for better visibility.